\documentclass[preprintnumbers,aps,prd,longbibliography,nofootinbib,nobibnotes,amsmath,amssymb]{revtex4}
\usepackage{graphicx,color,natbib}
\usepackage{amsmath,amssymb,amsfonts}
\usepackage{epsfig,epsf}
\usepackage{dcolumn}
\usepackage{float}
\usepackage{bm}
\usepackage{subfig}
\usepackage{caption}
\input amssym.def
\input amssym.tex

\usepackage[colorlinks=true, linkcolor=blue, bookmarks=true]{hyperref}

\begin{document}
\title{Entanglement of Purification as a Measure of Non-Conformality}
\author{M. Asadi\footnote{$\rm{m}_{-}$asadi@ipm.ir}}
\affiliation{School of Particles and Accelerators, Institute for Research in Fundamental Sciences (IPM), P.O.Box 19395-5531, Tehran, Iran}

\begin{abstract}
We have studied the entanglement of purification $E_p$ in a non-conformal holographic model which is a five-dimensional Einstein gravity coupled to a scalar field $\phi$ with a non-trivial potential $V(\phi)$. The dual 4-dimensional gauge theory is not conformal and exhibits a RG flow between two different fixed points. There are three parameters including energy scale $\Lambda$, model parameter $\phi_M$ and temperature $T$ which control the behaviour of the theory. Interestingly, we have found that $E_p$ can be used as a measure to probe the non-conformal behaviour of the theory at both zero and finite temperature. Furthermore, we have found that if one consider two different mixed state characterized by distinct values of $\frac{\Lambda}{T}$, then  the correlation between the subsystems of these states can be the same independent of $\frac{\Lambda}{T}$.
\end{abstract}
\maketitle

\tableofcontents
\section{Introduction}\label{sec:intro}

The gauge/gravity duality provides a significant framework to study key properties of the boundary field theory dual to some gravitational theory on the bulk side. The most concrete example of gauge/gravity duality is the AdS/CFT  correspondence which is a relationship between  a $d$-dimensional quantum field theory (QFT) to some classical gravitational theory in ($d+1$)-dimensions \cite{Maldacena:1997re,Witten:1998qj}. In fact, this correspondence is a strong-weak duality which is a novel approach to attack the problems of strongly coupled field theories using the classical gravity dual which is a weakly coupled theory. This provides an important framework to study the key properties of the boundary field theory dual to classical gravitational theory on the bulk side and has been utilized to study various areas of physics such as condensed matter, quantum information theory and the quark-gluon plasma \cite{Natsuume:2014sfa,CasalderreySolana:2011us,Hartnoll:2009sz,Camilo:2016kxq}. This duality provides also a key tool to investigate  physical quantities in the framework of the quantum information theory using their corresponding gravity duals such as entanglement entropy, mutual information and entanglement of purification. The key point is that it may be difficult to calculate these quantities in the field theory side, but one can compute them relatively simple  on the gravity side.

The generalization of gauge/gravity duality to field theories which are not conformal seems to be important and there are many examples of non-conformal field theories which one can particularly study different properties of them and then quantify the deviation of their observables from conformality through this duality. So, It is very interesting to study the effect of non-conformality on the physical quantities of the underlying theory \cite{Attems:2016ugt,Pang:2015lka,Rahimi:2016bbv,Asadi:2021nbd,Ali-Akbari:2021zsm,Lezgi:2020bkc,Asadi:2020gzl}.

In the context of quantum information theory the entanglement entropy $EE$ is one of the most important quantities which determines quantum entanglement between subsystems $A$ and its complementary $\bar{A}$  for a given pure state $\rho$. In the language of AdS/CFT correspondence, the entanglement entropy has a holographic dual given by the area of minimal surface extended in the bulk whose boundary coincides with the boundary
of the sub-system usually called Ryu and Takayanagi (RT)-surface \cite{Ryu:2006bv,Ryu:2006ef}.

 When we deal with a mixed state $\rho_{AB}$ which describes our total system, the $EE$ is not a appropriate quantity to measure the full correlation between two subsystems  $A$ and $B$. One of the most famous and finite quantity  to consider is the mutual information $I$ which measures the total correlation between subsystems  $A$ and $B$ \cite{Casini:2004bw,Wolf:2007tdq,Fischler:2012uv,Allais:2011ys,Hayden:2011ag,MohammadiMozaffar:2015wnx,Asadi:2018ijf,Ali-Akbari:2019zkf}.
   
  Another quantity which measures the total correlation between two disjoint subsystems $A$ and $B$ for mixed state $\rho_{AB}$ is  the entanglement of purification $E_p$ which  has recently received a lot of interest \cite{Terhal:2002riz}. It measures correlations (quantum and classical) between two disjoint subsystems $A$ and $B$ for a given mixed state described by density matrix $\rho_{AB}$. A candidate holographic counterpart for $E_p$  is the entanglement wedge cross section  $E_w$, the minimal cross section of the entanglement wedge \cite{Takayanagi:2017knl}.
 In order to understand various aspects of the $E_p$, many papers appear in the literature, for example, see \cite{Bhattacharyya:2018sbw,Espindola:2018ozt,Umemoto:2018jpc,Liu:2019qje,BabaeiVelni:2019pkw}.
  
The background we have considered in this paper is a five-dimensional Einstein gravity coupled to a scalar field with a non-trivial potential which has been introduced and studied in \cite{Attems:2016ugt}. In this model, the corresponding gauge theory is a non-conformal theory that has two conformal fixed points at UV and IR. In the gravity side, these solutions are asymptotically $\text{AdS}_5$ in the IR and UV limits with different radii. Thermodynamics transport and relaxation properties of the above background studied in \cite{Attems:2016ugt} and also this background utilized for studying entanglement entropy, subregion complexity and meson potential energy \cite{Rahimi:2016bbv,Asadi:2021nbd,Asadi:2020gzl}.

The remainder of this paper is organized as follows. Section \ref{sec:background} includes a brief review of the non-conformal model and its holographic dual in the gauge theory side at both zero and finite temperature. In section \ref{sec: Information measures} we represent a short but precise introduction of entanglement of entropy $EE$, mutual information $I$ and specifically entanglement of purification $E_p$. Section \ref{sec:Analytical} is devoted to compute $E_p$ holographically and analytically with a general asymptotically $AdS_{d+2}$ metric. Section \ref{sec:Results} is reserved for  study $E_p$ at both zero and finite temperature numerically. We will conclude in section \ref{sec:conclusion} with the discussion of our results. 

\section{Review on the background}\label{sec:background}
The non-conformal holographic model which we study here is a five-dimensional Einstein gravity coupled to a scalar field $\phi$ with a non-trivial potential $V(\phi)$ whose action is given by
\begin{align}\label{action}
S=\frac{2}{8 \pi G_5}\int d^5x\:\sqrt{-g}\:\big[\frac{1}{4}\mathcal{R}\:-\:\frac{1}{2}(\bigtriangledown\phi)^2\:-\:V(\phi)^2\big],
\end{align}
where $G_5$ is the five-dimensional Newton constant. In this paper we will consider a bottom-up model, which has been studied in detail in \cite{Attems:2016ugt},  and choose $V(\phi)$ to be negative which possess a maximum at $\phi=0$ and a minimum at $\phi=\phi_M>0$. Each of these extremum corresponds to an asymptotically $\text{AdS}_5$ solution whose radius is given by $L=\sqrt{\frac{-3}{V(\phi)}}$. We focus on the domain-wall solutions  interpolating between these two asymptotically $\text{AdS}_5$ solutions.

On the gauge theory side, each of these solutions is dual to a fixed point of the RG  flows from the UV fixed point at $\phi=0$ to the IR fixed point at $\phi=\phi_M>0$. The point is that if the potential $V(\phi)$ can be derived from a superpotential $W(\phi)$  as
 \begin{align}\label{VW}
 V(\phi)=\frac{-4}{3}W(\phi)^2+\frac{1}{2}W'(\phi)^2,
\end{align}
then, one can find the vacuum solutions to the Einstein equation by taking the metric of the following ansatz 
\begin{align}\label{metric}
ds^2=e^{2A(r)}(-dt^2+dx^2)+dr^2,\,\,\,\,\,\,\,\,\, \phi=\phi(r).
\end{align}
Inserting \eqref{metric} into the equations of motion we find that $A(r)$ and $\phi(r)$ satisfy the first order equations
\begin{align}\label{first order eq}
\frac{dA}{dr}=-\frac{2}{3}W\,\,,\,\,\,\,\,\,\,\frac{d\phi}{dr}=\frac{dW}{d\phi}.
\end{align}
By choosing  the following superpotential which is characterized by a single parameter $\phi_M$, which we call it the model parameter since then, \cite{Attems:2016ugt} 
\begin{align}\label{superpotential}
LW(\phi)=-\frac{3}{2}-\frac{\phi^2}{2}+\frac{\phi^4}{4\phi_M^2},
\end{align}
and using eq.\eqref{VW}, the potential is given by
\begin{align}\label{potential}
L^2V(\phi)=-3-\frac{3}{2}\phi^2 -\frac{1}{3}\phi^4 + \big(\frac{1}{3\phi_M^2 }+\frac{1}{2\phi_M^4}\big)\phi^6 -\frac{1}{12\phi_M^4}\phi^8.
\end{align}
Indeed, this potential have a maximum at $\phi=0$ whose resulting geometry is asymptotically $\text{AdS}_5$ in the  UV  and a minimum at $\phi=\phi_M>0$ where the resulting geometry near  it is  again asymptotically $\text{AdS}_5$ in the  IR. It is easily seen that the radii of these solutions take the form
\begin{eqnarray}\label{Lads}
{L=\sqrt{\frac{-3}{V(\phi)}}}=
\begin{cases}
L_{\text{UV}}=L & \: \phi=0 , \\
L_{\text{IR}}=\frac{L}{1+\frac{1}{6\phi_M^2}} & \: \phi=\phi_M , \\
\end{cases}
\end{eqnarray}
where it is obvious that $L_{\text{IR}}<L_{\text{UV}}$. There are two important points which are needed to mention here. First, according to gauge/gravity dictionary the number of degrees of freedom $N$ is proportional to  $L$ through the relation $N^2\propto \frac{L^3}{G_5}$  which together with eq. \eqref{Lads} yields the fact that as we move from an UV to an IR fixed point the number of degrees of freedom decreases. Second, if we increase $\phi_M$, then the difference in degrees of freedom between the UV and the IR fixed points, which we call it $\Delta N$, will increase. 

Choosing superpotential eq. \eqref{superpotential} and solving eq. \eqref{first order eq}, one can obtain analytically the vacuum solution for arbitrary model parameter $\phi_M$ as follows\cite{Attems:2016ugt} 
\begin{align}
e^{2A(r)}&=\frac{\Lambda ^2 L^2}{\phi(r) ^2}(1-\frac{\phi(r) ^2}{\phi _M ^2})^{1+\frac{\phi _M^2}{6}}\,\,\,\,\, e^{-\frac{\phi (r)^2}{6}},\\
\phi (r)&=\frac{\Lambda L e^{-\frac{r}{L}}}{\sqrt{1+\frac{\Lambda ^2L^2}{\phi _M^2}e^{-\frac{2r}{L}}}},
\end{align}
where $\Lambda$ is the energy scale that break the conformal symmetry explicitly. In the dual gauge theory side, $\Lambda$ is identified with the source of the dimension$-3$ operator $\mathcal{O}$.\\
It is noticed that $\phi(r\rightarrow\infty)=0$ which confirms that $\phi=0$ is a solution, while $\phi(r=0)=\frac{L\Lambda}{\sqrt{1+\frac{L^2\Lambda^2}{\phi_M^2}}}$ and hence  in order to have $\phi=\phi_M$ as a solution we need to consider $\frac{\phi_M^2}{L^2\Lambda^2}\ll 1$. The same argument should be checked for $e^{2A (r)}$. With the radial variable $r$ in \eqref{metric} the asymptotic $AdS$ boundary region is $r\rightarrow\infty $, so we expect $e^\frac{2A (r)}{L}\rightarrow e^\frac{2r}{L_{\text{UV}}}$ in this region. The same behaviour is going to be valid at $r=0 $ with condition $\frac{\phi_M^2}{L^2\Lambda^2}\ll 1$, i.e. $e^\frac{2A (r)}{L}\rightarrow e^\frac{2r}{L_{\text{IR}}}$,   which are both verified in those two regions  so that eq. \eqref{Lads} is satisfied. To summarized, the vacuum solution describes a RG flow from an UV fixed point, $\phi(r\rightarrow\infty)=0$,  to an IR fixed point, $\phi(r=0)=\phi_M\,\, \text{with}\,\,\frac{\phi_M^2}{L^2\Lambda^2}\ll 1$.

Now we would like to study the thermal physics of the non-conformal model described in \eqref{action} whose background is given by the following metric
 \begin{eqnarray}\label{metric}
ds^2=e^{2A(\phi)}(-h(\phi)dt^2+d \vec{x} ^2)+\frac{e^{2B(\phi)}}{h(\phi)}d\phi^2,
\end{eqnarray}
where $A$, $B$, and $h$ are functions of $\phi$, and $\phi$ is also some function of $r$. There is a horizon at $\phi=\phi_H$ which is the solution to the equation $h(\phi_H)=0$ and the interval $0<\phi<\phi_H$ corresponds to the outside of the horizon. We assume that $A(\phi)$ and $B(\phi)$ are finite at the horizon . For later convenience, we express the above metric in Eddington-Finkelstein form

  \begin{eqnarray}\label{metric2}
ds^2=e^{2A(\phi)}(-h(\phi)d\tau ^2+d \vec{x} ^2)- 2e^{A(\phi)+B(\phi)} L\,d\tau d\phi.
\end{eqnarray}
One can find a black hole solution if a master function $G(\phi)$, where $G(\phi)=A'(\phi)$, is defined \cite{Gubser:2008ny}. Using  $G(\phi)$ and knowing $V(\phi)$ the different metric components are given by Einstein's equations  \cite{Attems:2016ugt}
\begin{subequations}\label{metric3}
\begin{align}
A(\phi)&=-\log\left(\frac{\phi}{\phi_0}\right)+\int_{0}^{\phi}d\tilde{\phi}\left(G(\tilde{\phi})+\frac{1}{\tilde{\phi}} \right),  \\
B(\phi)&=\log\left(|G(\phi)|\right) +\int_{0}^{\phi}d\tilde{\phi}\frac{2}{3G(\tilde{\phi})},\\
h(\phi)&=-\frac{e^{2B(\phi)}L^2\left(4V(\phi)+3G(\phi)V'(\phi) \right) }{3G'(\phi)},
 \end{align}
\end{subequations}
where $G(\phi)$ must satisfy the following non-linear master equation
\begin{align}\label{a}
\frac{G'(\phi)}{G(\phi)+\frac{4V(\phi)}{3V'(\phi)}}&=\frac{d}{d\phi}\log\bigg[\frac{1}{3G(\phi)}-2G(\phi) 
+\frac{G'(\phi)}{2G(\phi)}-\frac{G'(\phi)}{2\left(G(\phi)+\frac{4V(\phi)}{3V'(\phi)}\right)}\bigg]. 
\end{align}
From these metric coefficients and their relation with the master function, one can obtain the following expression for the Hawking temperature T
\begin{align} \label{temperature}
\frac{T}{\Lambda}=- \frac{L^2  V(\phi_H)}{3 \pi \phi_H} \exp \left[ \int_0^{\phi_H} d\phi \left(    G(\phi) +\frac{1}{\phi}+ \frac{2}{3 G(\phi)}\right) \right].
\end{align}

 \section{Brief review on the entanglement entropy, mutual information and Entanglement of purification }\label{sec: Information measures}
\begin{figure}
\centering
\includegraphics[width=120 mm]{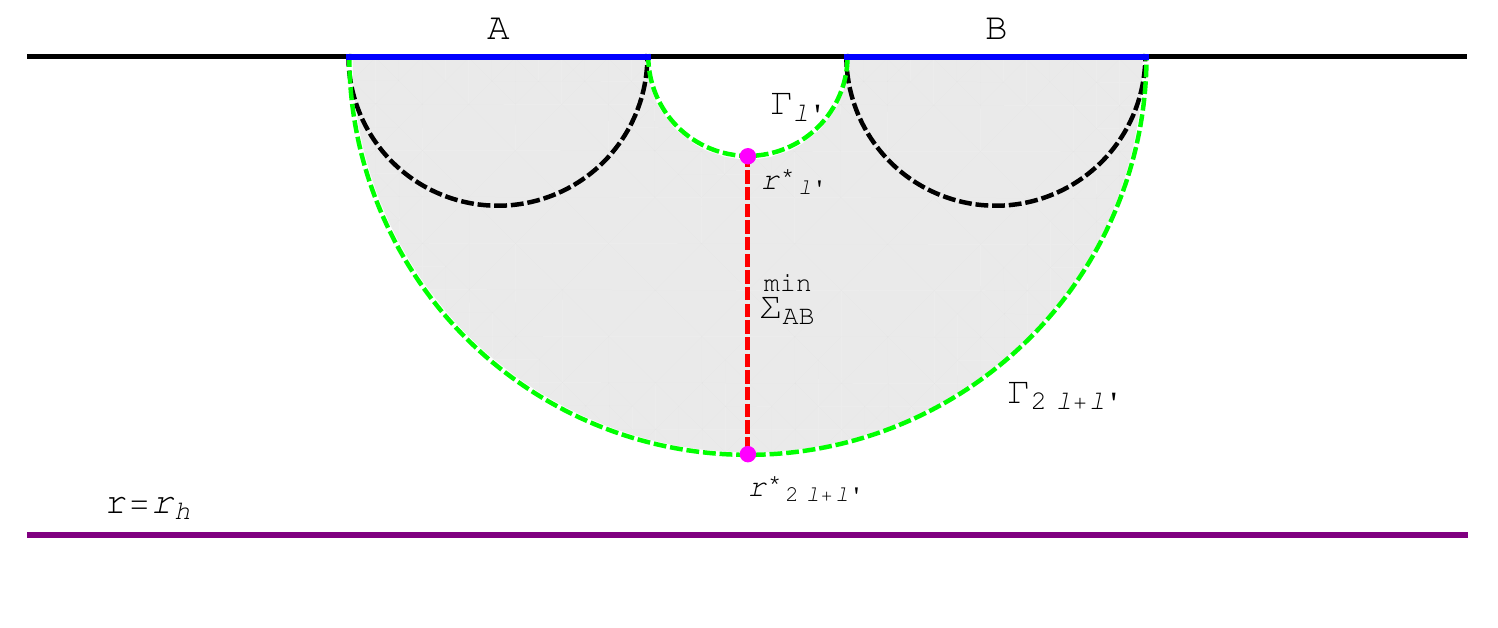}
\caption{The gray region shows the entanglement wedge dual to $\rho_{AB}$. The minimal surfaces, RT-surfaces, are denoted by $\Gamma$, the dashed curves.}
\label{Eop1}
\end{figure}

 \begin{itemize}
 \item Entanglement entropy
 
 Consider a constant time slice in a $d$-dimensional quantum field theory which is described by the pure state $|\psi\rangle$  and density matrix $|\psi\rangle\langle\psi|$  and divide it into two spatial regions $A$ and its complement $\bar{A}$. Then, the total Hilbert space is factorized into 
 $\cal{H}_{\textbf{tot}}= {\cal{H}}_{A}\otimes{\cal{H}}_{\bar{A}}$. The reduced density matrix for region $A$ can be calculated by integrating out the degrees of freedom living in $\bar{A}$, i.e. $\rho_A=Tr_{\bar{A}}\,\rho$. The entanglement between subregions $A$ and $\bar{A}$ is measured by the entanglement entropy which is a non-local quantity and defined as the Von-Neumann entropy of the reduced density
 \begin{align}
S_{A}=-tr \rho_{A}\log\rho_{A}.
\end{align}
 A holographic prescription, known as Ryu and Takayanagi ($RT$) proposal, has been proposed to compute entanglement entropy through the following area law relation\cite{Ryu:2006bv,Ryu:2006ef}
 \begin{align}\label{RT}
S_{A}=\frac{Area(\gamma_{A})}{4G_{N}^{d+2}},
\end{align}
where $S_{A}$ is the holographic entanglement entropy for the subregion $A$, $\gamma_{A}$  is a codimension-two minimal area surface in the bulk ($RT$-surface), whose boundary $\partial \gamma_{A}$ coincides with $\partial A$, and $G_{N}^{d+2}$  is the $d+2-$ dimensional Newton constant.

 \item Mutual information
 
 When the boundary entangling region is made by two disjoint subregions, an important quantity to study is the mutual information $I$ which is a quantity  derived from entanglement entropy whose definition is given by \cite{Fischler:2012uv}
 \begin{align}
I(A,B)=S_A+S_B-S_{A\cup B},
\end{align}
 where $S_A$, $S_B$ and $S_{A\cup B}$ denote the entanglement entropy of the region $A$, $B$ and $A\cup B$, respectively and can be calculated by RT-prescription. The mutual information $I$  is a finite, positive, semi-definite quantity which measures the total correlation between the two subregions $A$ and $B$ \cite{Groisman:2005dbo}.
 \item Entanglement of purification
 
 The entanglement of purification  is an important quantity which measure the total (quantum and classical) correlation between two disjoint subregions for a given mixed state \cite{Bhattacharyya:2019tsi,Terhal:2002riz}. Consider a bipartite system described by a mixed state and density matrix $\rho_{AB}$. We can always purify this mixed state by enlarging its Hilbert space as $\mathcal{H}_A\otimes \mathcal{H}_B\rightarrow \mathcal{H}_A\otimes \mathcal{H}_B \otimes \mathcal{H}_{A'}\otimes \mathcal{H}_{B'}$ such that the total density matrix in enlarged Hilbert space $\rho_{AA'BB'}$ is given by $\rho_{AA'BB'}=\vert \psi_{AA'BB'}\rangle\langle\psi_{AA'BB'}\vert$. Such a pure state is called a purification of $\rho_{AB}$ if we have
\begin{align}
\rho_{AB}=Tr_{A'B'}\left(\vert \psi_{AA'BB'}\rangle\langle\psi_{AA'BB'}\vert\right).
\end{align}

Obviously there exist infinite ways to purify $\rho_{AB}$. The EoP is defined by minimizing the entanglement entropy $S_{AA'}$ over all purifications of $\rho_{AB}$ \cite{Terhal:2002riz}
\begin{align}
E_p(\rho_{AB})=\underset{\vert \psi _{AA'BB'}\rangle}{\rm{min}}(S_{AA'}),
\end{align}
where $S_{AA'}$ is the entanglement entropy corresponding to the density matrix $\rho_{AA'}$ and $\rho_{AA'}={\rm{Tr}}_{BB'}\big[\left(|\psi\rangle_{ABA'B'}\right) \left({}_{ABA'B'}\langle\psi|\right)\big]$.

In the context of the gauge/gravity duality, it has been conjectured  that the $E_p$ is dual to the entanglement wedge cross-section $E_w$ of $\rho_{AB}$ which is defined by \cite{Takayanagi:2017knl,Nguyen:2017yqw}
\begin{align}\label{EWCS}
E_w(\rho_{AB})=\frac{{\rm{Area}}(\Sigma_{AB}^{min})}{4G_N^{(d+2)}}.
\end{align}
where $\Sigma_{AB}^{min}$ is the minimal surface in the entanglement wedge $E_w(\rho_{AB})$, separating two subregions of A and B, that ends on the RT-surface of $A\cup B$, the Red-dashed line in Fig. \ref{Eop1} and, as a result, it is conjecture that \cite{Takayanagi:2017knl}
\begin{align}\label{eop}
E_w(\rho_{AB})= E_p(\rho_{AB}).
\end{align}
 \end{itemize}

 \section{Analytical prescription}\label{sec:Analytical}
In this subsection we would like to compute holographic entanglement of purification $E_p$ analytically. Therefore, we start with a general asymptotically $AdS_{d+2}$ metric 
\begin{align} \label{generalmetric}
ds^2=f_1(r)dt^2+f_2(r)dr^2+f_3(r)d\vec{x}^2,
\end{align}
where $r$ is the radial direction and $\vec{x}\equiv (x_1,...,x_d)$.  The strongly coupled field theory lives on the boundary at $r\rightarrow \infty$ and $f_1$, $f_2$ and $f_3$ are arbitrary functions of $r$ which will be fixed later on.  We consider a symmetric configuration of two disjoint subregions $A$ and $B$  at a constant time slice. More specifically, $A$ and $B$ are two parallel strips with equal widths $l$ extended along $x_i$ direction with length $L$ and separated by a distance $l'$ in the $x_1$ direction. This configuration is given by the following profile, $x(r)$, which is an even function of $r$, see Fig. \ref{Eop1}
\begin{align}
\begin{split} %
x(r)& \equiv x_1(r)\cr
-\frac{L}{2}&\leq x_i \leq \frac{L}{2},\ \ \ i=2,...,d,\,\,\,\,L\rightarrow\infty .
\end{split}
\end{align}

 It is obvious that the minimal surface in the entanglement wedge, which is $\Sigma_{AB}^{min}$, runs along the radial direction $r$ and connects the turning points of minimal surfaces $\Gamma_{l'}$ and $\Gamma_{l'+2l}$. Now, using  \eqref{EWCS} and \eqref{generalmetric} holographic entanglement of purification $E_p$ is given by 
\begin{align}\label{heop}
E_w=\frac{L^{d-1}}{4G_N}\int_{r^*_{2l+l'}}^{r^*_{l'}}dr \sqrt{f_2 f_3^{d-1}},
\end{align}
where $r^*_{l'}$ and $r^*_{2l+l'}$ denote the turning point of $\Gamma_{l'}$ and $\Gamma_{2l+l'}$, respectively. If we would like to find the turning point $r^*_{l'}$, then we will need to consider the following configuration
\begin{align}
\begin{split} %
-\frac{l'}{2}&\leq x(r) \equiv x_1(r)\leq \frac{l'}{2},\cr
-\frac{L}{2}&\leq x_i \leq \frac{L}{2},\ \ \ i=2,...,d,
\end{split}
\end{align}
whose area is written as
\begin{align}\label{area}
{\rm{Area}}=2L^{d-1}\int_{r^*_{l'}}^\infty  dr f_3^{\frac{d-1}{2}} \sqrt{f_2-f_3\ x'(r)^2}\equiv \int dr {\cal{L}}.
\end{align}
It is clearly seen from the above expression that ${\cal{L}}$ does not depend on $x$ explicitly and hence the corresponding Hamiltonian is constant. Therefore, it is then easy to find $r^*_{l'}$ as a function of $l'$
\begin{align}\label{length} %
\frac{l'}{2}=\int_{r^*_{l'}}^\infty dr \sqrt{\frac{f_2f_{3*}^d}{f_3(f_{3}^d-f_{3*}^d)}},
\end{align}

where $f_{\ast}\equiv f(r^*_{l'})$. Note that the above equation gives the relation between $l'$ and $r^*_{l'}$. A similar equation for $r^*_{2l+l'}$ is achieved by replacing  $l'$ and $r^*_{l'}$ with $2l+l'$ and $r^*_{2l+l'}$.

\section{Numerical result}\label{sec:Results}
Having set up the general framework and ingredient of  holographic entanglement of purification $E_p$, now we are ready to study the numerical results of our model for both zero and finite temperature cases. To do so, we consider two subregions $A$ and $B$ with the same length $l$ which is separated by length $l'$.
\begin{figure}
\centering
\includegraphics[scale=0.38]{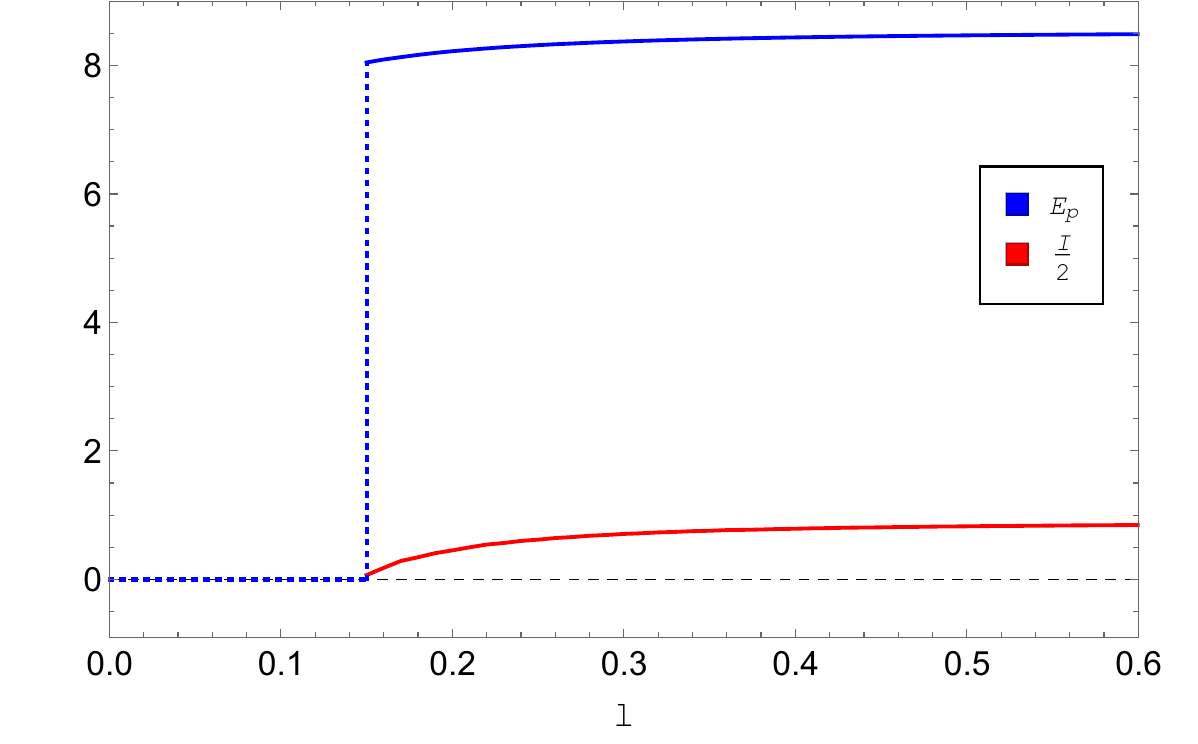}
\includegraphics[scale=0.38]{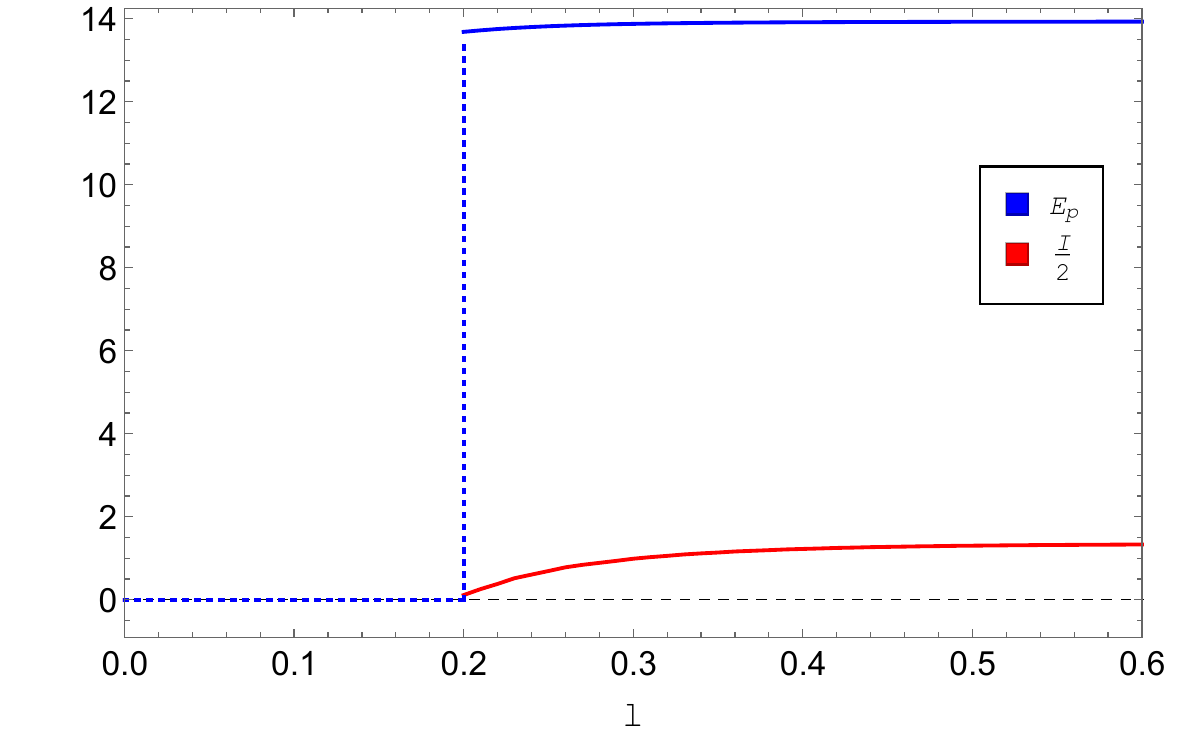}
\caption{$E_p$ and $I/2$  with rspect to $l$ for $l'=0.11$ (left) and $l'=0.2$ (right). We set $\Lambda=20$ and $\phi_M=2$.}\label{Eop2}
\end{figure}
\begin{figure}
\centering
\includegraphics[scale=0.42]{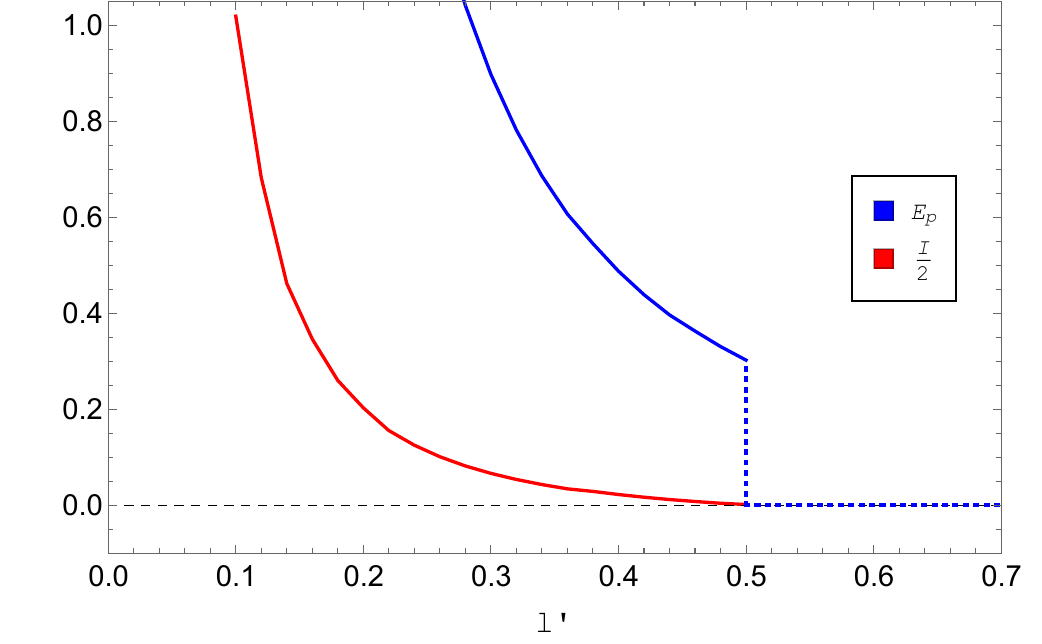}
\includegraphics[scale=0.42]{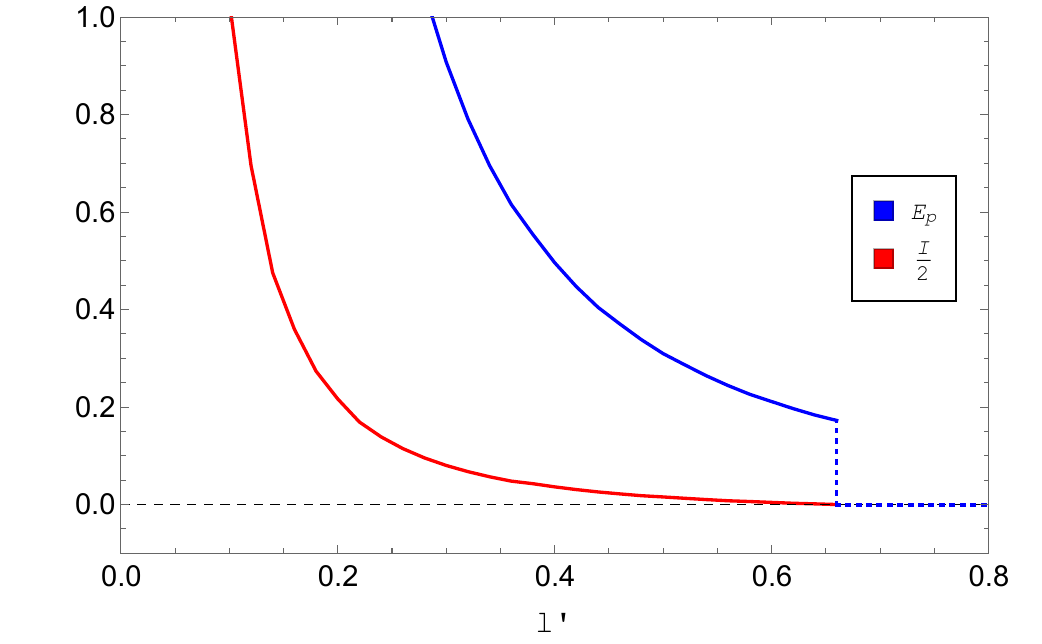}
\caption{$E_p$ and $I/2$  with rspect to $l'$ for $l=0.7$ (left) and $l=0.9$ (right). We set $\Lambda=20$ and $\phi_M=2$.}\label{Eop3}
\end{figure} 
 
\subsection{Zero temperature}\label{sec:Zero}

\begin{figure}
\includegraphics[scale=0.33]{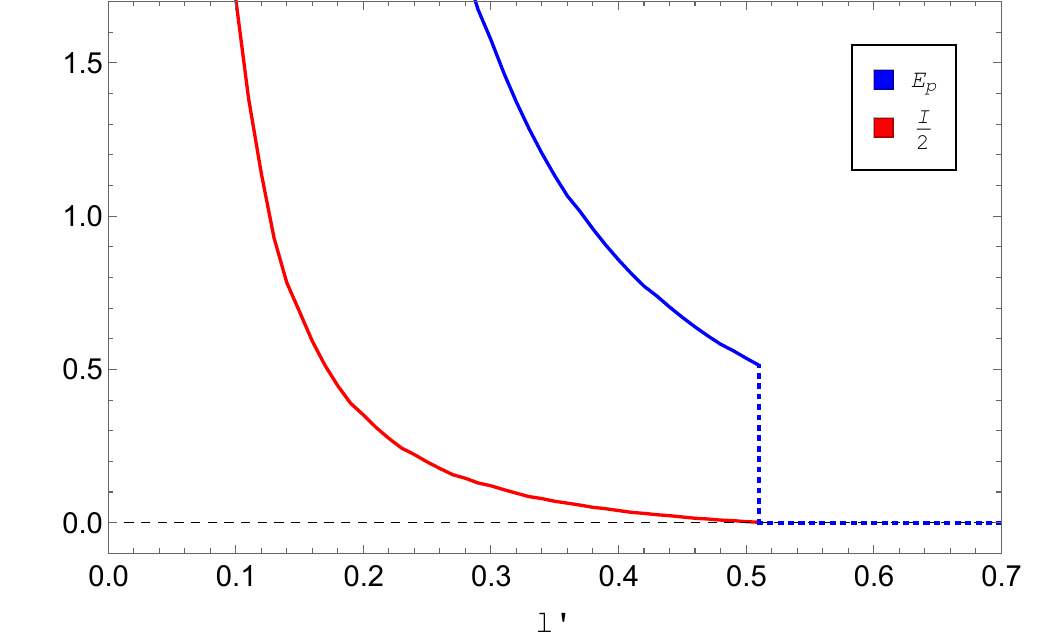}
\includegraphics[scale=0.33]{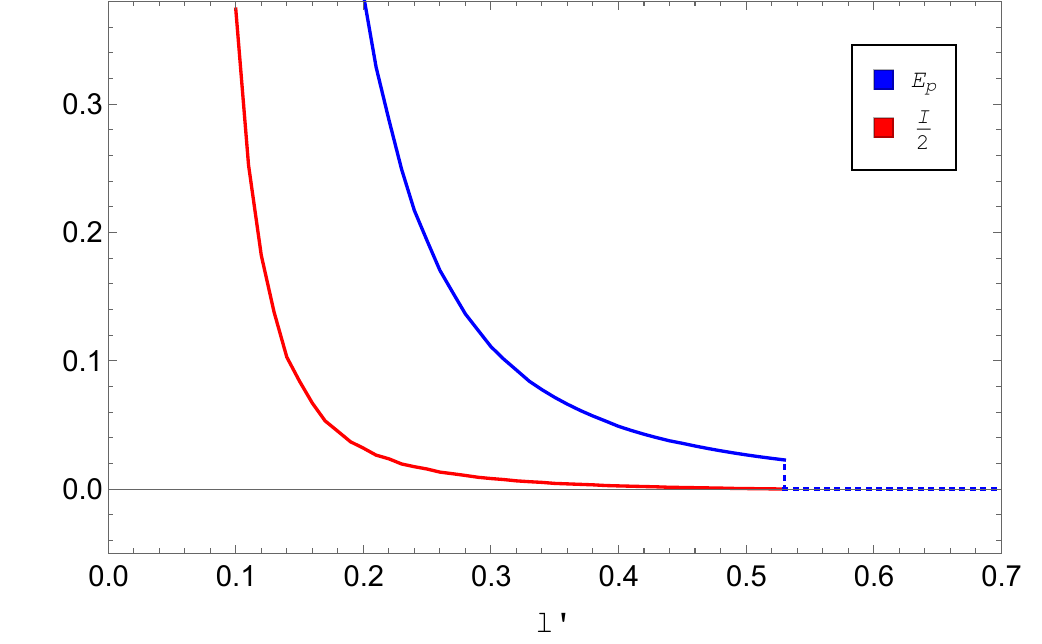}
\includegraphics[scale=0.33]{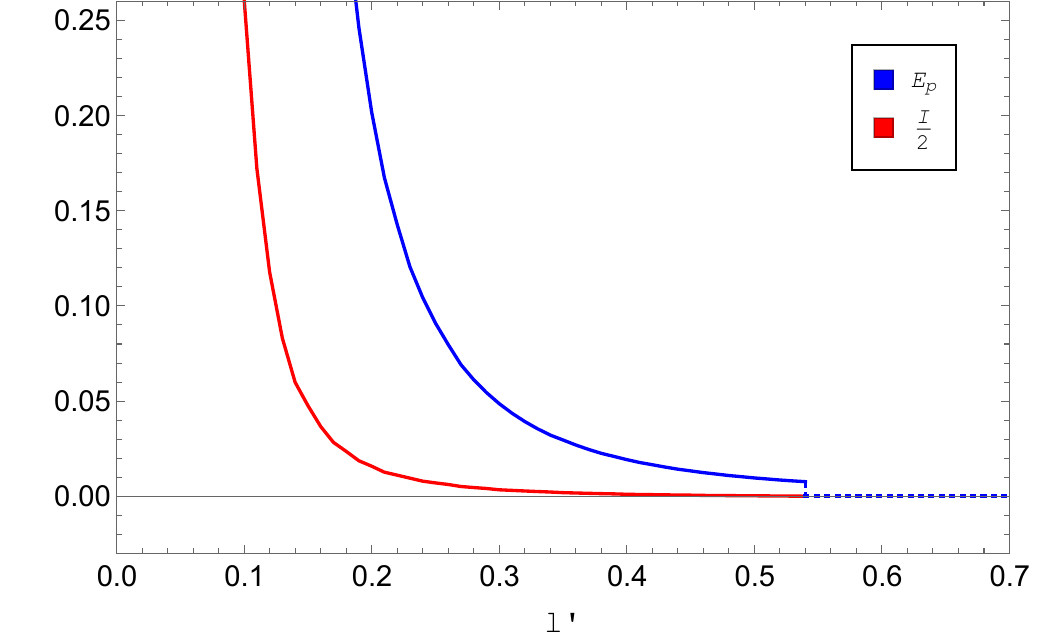}
\caption{$E_p$ and $I/2$  with rspect to $l'$ for $l=0.7$ and we set $\Lambda=15$ for different  $\phi_M$. Left: $\phi_M=1.5$. Middle: $\phi_M=5$. Right: $\phi_M=10$ (right).}\label{Eop4}
\end{figure}
\begin{figure}
\centering
\includegraphics[scale=0.5]{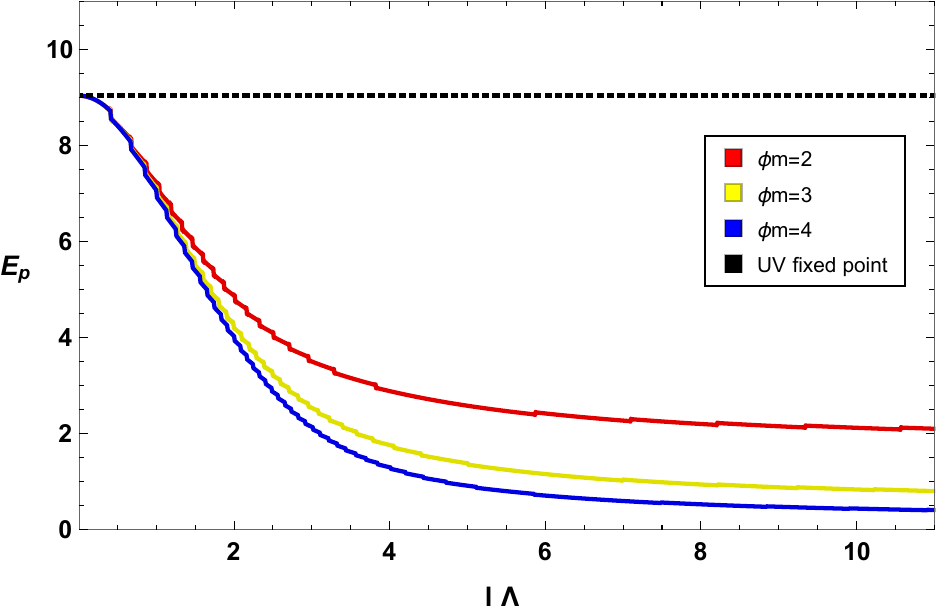}
\caption{$E_p$  with rspect to $l\Lambda$ for $l=0.5$ and $l'=0.2$. Different curves correspond to distinct  $\phi_M=2$ (red), $\phi_M=3$ (blue), $\phi_M=4$ (green).The black dashed curve is the value of $E_p$ at the UV fixed point.}\label{Eop5}
\end{figure}
In Fig.\ref{Eop2} we have plotted  $E_p$ and $I/2$  with respect to $l$ . We fix $l'=0.11$ (left panel), $l'=0.2$ (right panel) and set $\Lambda=20$ and $\phi_M=2$. As it is evident, the common feature is that there is a specific length $l$  before  which $E_p$ and $I/2$ are zero which means there is no correlation between two subregions. In the bulk view, the $RT$-surface for $S_{A\cup B}$ disconnects resulting in $E_p=I/=0$.  Then, a phase transition of $S_{A\cup B}$ to the connected phase happens which is marked by the point where both $E_p$ and $I/2$ become zero. The point is that $I$ becomes zero continuously while $E_p$ jumps discontinuously at the phase transition.  At the end they monotonically and slowly increase   to a saturated value.
Different values of $l'$ yield qualitatively similar curves: monotonically increasing curves approach to a constant value.
 As we expect for large enough values of $l$ one can say that the turning point $r^*$ approaches asymptotically $\text{AdS}_5$ in the IR limit which is conformal and therefore the $E_p$ and $I/2$ goes to a constant value. In Fig. \ref{Eop3}, we plot $E_p$ and $I/2$  with respect to $l'$. We fix $l=0.2$ (left panel), $l=0.9$ (right panel) and set $\Lambda=20$ and $\phi_M=2$. This figure also shares the same characteristics as Fig.\ref{Eop2}. We again observe the phase transition from connected configuration to disconnected one of $S_{A\cup B}$ for specific length $l'$ which we call it the disentangling length $l'_D$ for later convenience. Another important point is that there is an inequality between  $E_p$ and $I/2$ which has been proven in holography \cite{Takayanagi:2017knl} and is written as
 \begin{align}
 \frac{I}{2}\leq E_p\, ,
 \end{align}
 where we have checked the above inequality in both Figs. \ref{Eop2} and \ref{Eop3}  and find that it is indeed satisfied.

In Fig. \ref{Eop4}, we have sketched Fig. \ref{Eop3} for different model parameter $\phi_M$ to investigate how this model parameter  affect the disentangling length $l'_D$.
It is observed that if we  increase $\phi_M$, which yields the difference in degrees of freedom between the UV and the IR fixed points $\Delta N$ increases, then $l'_D$ will become longer, $(l'_D)_{\phi_M=1.5}<(l'_D)_{\phi_M=5}<(l'_D)_{\phi_M=10}$. In other words, by rising $\Delta N$ the correlation between two subregions will be vanished at longer separation length.

In Fig. \ref{Eop5}, $E_p$ with respect to $l\Lambda$ for $l=0.5$ and $l'=0.2$ and different model parameter $\phi_M=2$ (red), $\phi_M=3$ (blue) and $\phi_M=4$ (green) has been plotted. The black dashed curve corresponds to the value of $E_p$ at the UV fixed point. The interesting point is that 
$E_p$ exhibits moderate change due to the non-conformality for large $\Lambda$ that is the larger model parameter $\phi_M$, which is equivalent to larger $\Delta N$, the larger the deviations from conformality in the theory. Hence, we can conclude that $E_p$ is a good measure to quantify the deviation from the conformality. While, for small enough $\Lambda$ we observe that $E_p$ is independent of $\Lambda$. There are also some  common, but interesting, features which we would like to mention here. The first one is that the different curves start from the value of $E_p$ at the UV fixed point, then experience a slowly-decreasing stage at small enough $\Lambda$ along with a quickly-reduction stage at intermediate $\Lambda$, and at the end saturate to  the value of $E_p$ at their IR fixed point. Sufficiently deep in the IR, It is also observed that $E_p$ approaches its asymptotic value and we can say that the system behaves again as approximately conformal. However, the model parameter $\phi_M$ plays the key role to determine at which scale this can be happened. Another point is that as $\phi_M $ increases, $E_p$ approaches its asymptotic value more slowly. 
\subsection{Finite temperature}\label{sec:finitT}
\begin{figure}
\centering
\includegraphics[scale=0.42]{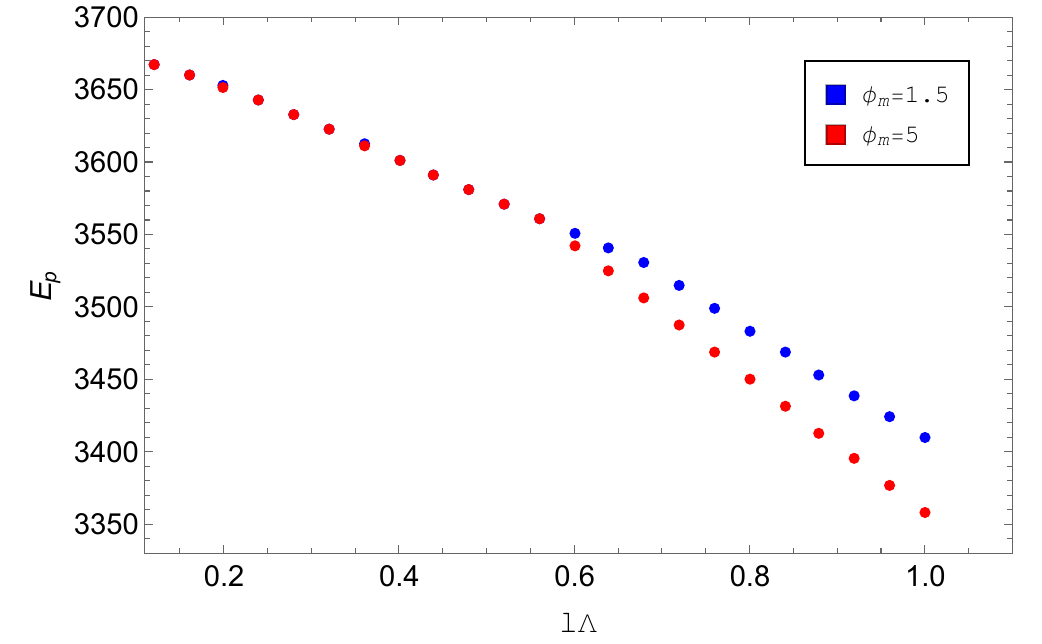}
\includegraphics[scale=0.42]{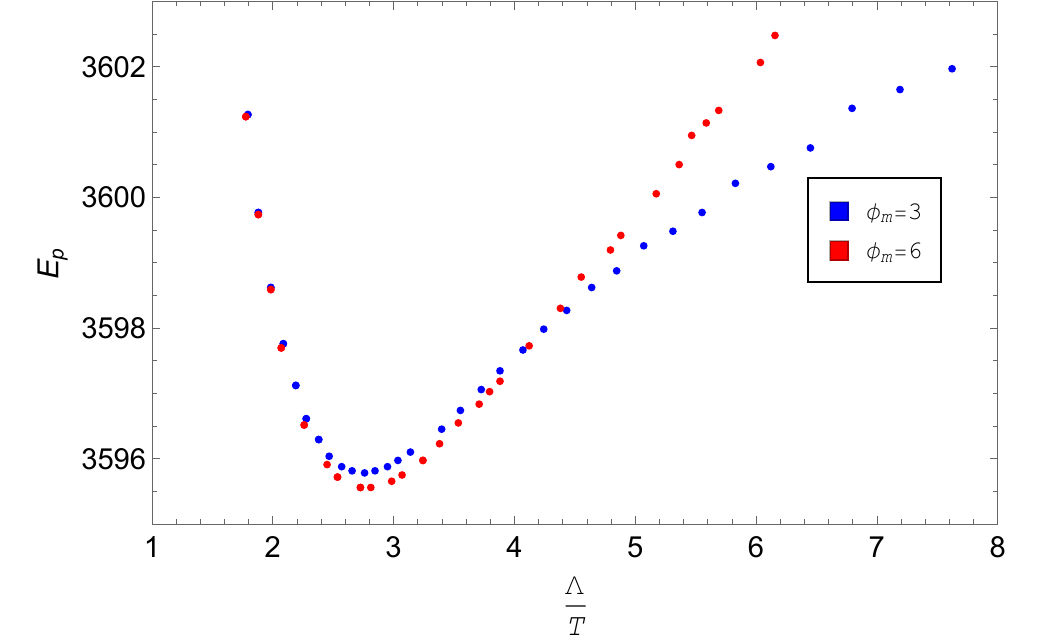}
\caption{Left: $E_p$  with rspect to $l\Lambda$ for $l=0.04$ and $l'=0.01$. Different curves correspond to distinct  $\phi_M=1.5$ (blue) and $\phi_M=5$ (red). Right:  $E_p$  with rspect to $\frac{\Lambda}{T}$ for $l=0.04$ and $l'=0.01$. We set $\Lambda=10$ and $\phi_M=3$. We set $\Lambda=20$ and $\phi_M=2$.}\label{Eop6}
\end{figure}

In the left panel of Fig. \ref{Eop6}, $E_p$ with respect $l\Lambda$ for $l=0.04$ and $l'=0.01$ and different model parameter $\phi_M=1.5$ (blue) and $\phi_M=5$ (red)  has been plotted. It is seen that by increasing energy scale $\Lambda$  the entanglement of purification $E_p$  decreases which is the same behaviour as we find at zero temperature. It is also observed that if we increase  $\phi_M $, which results in raising the difference in degrees of freedom between the UV and IR fixed points $\Delta N$, $E_p$  becomes smaller. For small enough $\Lambda$ the two curves with different $\phi_M $ coincide with each other and hence $E_p$ is independent of model parameter $\phi_M $. While, for large $\Lambda$, one can observe that  the larger  $\phi_M $ which is corresponds to the bigger $\Delta N$,  the larger the deviations from conformality in the theory and we can deduce that $E_p$ is a appropriate measure of the non-conformality of the theory at finite temperature. 
In the right panel of  Fig. \ref{Eop6} we have plotted $E_p$ as a function of $\frac{\Lambda}{T}$ for fixed $l=0.04$ and $l'=0.01$.  We set $\Lambda=10$ and two different model parameter $\phi_M=3$ and  $\phi_M=6$ have been considered. It is seen that $E_p$ is not a monotonic function with respect to $\frac{\Lambda}{T}$ starting from a definite positive  value and decreases to a minimum value with a steep slope and finally experience a gradually growth. Since  $E_p$ is a measure of total correlation between two subsystems, accordingly, the thermal fluctuations can promote both entangling and disentangling between them. As mentioned, there is a minimum value for $E_p$ where it is a monotonically decreasing function at high temperature ($T\gg\Lambda$)  and intermediate temperature ($T\sim\Lambda$) regime but it is a monotonically increasing function at low temperature ($T\ll\Lambda$)  regime. This feature interestingly suggest that there exist two values of $\frac{\Lambda}{T}$ which have the same value of $E_p$ and hence if we consider  different mixed states characterized by distinct  $\frac{\Lambda}{T}$, then the correlation between the subsystems of those states can be the same independent of $\frac{\Lambda}{T}$. This behaviour has also been reported by the authors in a model which enjoys a critical point in its phase space \cite{Amrahi:2020jqg}. Another interesting point is that, at  low temperature ($T\ll\Lambda$)  regime, $E_p$ is totally $\phi_M$ dependent and consequently, we can use $E_p$  as a measure of the non-conformality of the theory at finite temperature. It is also seen that  $E_p$ is independent of model parameter $\phi_M$ at high temperature ($T\gg\Lambda$)   regime  . This is due to the fact that the value of the scalar field at horizon is  small and hence the physics is sensitive only to the small field behaviour of the scalar potential which is independent of $\phi_M$. 
\section{Conclusion}\label{sec:conclusion}
In this paper we have explored the entanglement of purification $E_p$  for a non-conformal holographic model at both zero and finite temperature. The model is a five-dimensional Einstein gravity coupled to a scalar field $\phi$ with a non-trivial potential $V(\phi)$ and we consider  two parallel strips with the same length, which are symmetric configurations. We have sought to answer the question of whether $E_p$  can be used as a measure to probe the non-conformal property of our model and the results have shown that the answer is yes. Indeed, we have found that  at both zero and finite temperature $E_p$ is a good measure to probe the non-conformality of the theory. Furthermore, $E_p$ is not a monotonic function with respect to $\frac{\Lambda}{T}$ at finite temperature. It could be an increasing or decreasing function of $\frac{\Lambda}{T}$ depending on which temperature  region we look at. In other words, suppose we are given two different mixed states labeling with distinct value of $\frac{\Lambda}{T}$, then $E_p$ says that these states  have subsystems whose correlation is the same independent of $\frac{\Lambda}{T}$.

\section*{Acknowledgement} 

We would like to thank Mohsen Alishahiha for fruitful discussions and his supports. This work is based upon research funded by Iran National Science Foundation (INSF) under project No. 4024022.


\end{document}